\documentclass[amsmath,amssymb,aps,prfluids]{revtex4-2}
\usepackage{tikz}
\usepackage{amsmath}
\usetikzlibrary{arrows.meta,calc,angles,quotes,decorations.markings}
\usepackage{enumerate}
\usepackage{soul}
\newcommand{\mr}{m_{\text{rider}}}
\newcommand{\mf}{m_{\text{foil}}}

\begin{document}


\title{A minimal model of pump foil dynamics} 


\author{Eunok Yim$^1$ and Francois Gallaire$^2$}
 \affiliation{$^1$HEAD-Lab., EPFL,  Lausanne, Switzerland \\
 $^2$LFMI, EPFL,  Lausanne, Switzerland }
 


\date{\today}
\begin{abstract}
Pump foiling enables a hydrofoil surfboard to sustain forward motion on flat water using only periodic leg pumping, converting vertical oscillations into hydrodynamic lift and thrust. We present a minimal mechanical model of pump foil propulsion, formulated as a coupled second-order system for horizontal and vertical translation and pitch in an inertial frame. The rider-board system is modeled in reduced form, with the rider mass concentrated at the body center and the foil mass assigned to the mast-linker pivot, about which the front and rear wing dynamics are written. Hydrodynamic loading includes quasi-steady lift and drag, buoyancy, and rotational effects, including rotational lift and nonlinear rotational drag. Under simplified control assumptions in which the pumping frequency is fixed and the net rider pumping input is represented by a single effective force-amplitude parameter, the model predicts sustained stable forward-propulsion regimes at modest forcing amplitude and small positive pitch angle. In this regime, the front wing acts primarily as the main lifting surface, while the rear wing, although contributing less to vertical support, is essential for pitch stability because of its longer moment arm. These results provide a mechanistic interpretation of pump foil propulsion and identify measurable quantities and parameter sensitivities that can guide targeted field and laboratory experiments and help refine assumptions on rider control inputs and hydrodynamic loading.\end{abstract}


\maketitle

\section{Introduction \label{sec:Intro}}

Pumped hydrofoils have recently emerged as a novel recreational water sports, where sustained forward motion of a waterborne vehicle is achieved through periodic vertical pumping by a rider rather than by an external propulsion system. In this configuration, a board is elevated above the free surface and connected via a mast to one or more submerged hydrofoils. Despite the apparent absence of buoyancy or mechanical propulsion, the system is capable of maintaining lift and forward speed, a behavior that may appear counterintuitive to an external observer.


As noted in  \cite{rozhdestvensky2023simplified}, pump foiling provides a compelling example of unsteady hydrofoil propulsion in a near surface environment. Through rhythmic vertical motion of the coupled hydrofoil-board system, the rider simultaneously generates lift and thrust. The problem therefore belongs to the broader class of water sports hydrodynamics involving the generation of unsteady forces (see, e.g., \cite{Labbe2019, Pretot2022}). Accordingly, several theoretical and experimental studies have examined this unsteady propulsion mechanism by deriving analytical expressions for thrust and lift in the potential-flow limit, highlighting the roles of unsteady circulation and vortex dynamics during heaving and pitching cycles (see \cite{AlaminosQuesada2022} for a recent contribution).
More recent studies have started to investigate the fully coupled interactions between oscillating foils and ambient surface waves, showing that submergence depth and phase relations with incident wave fields can modify propulsive performance and efficiency in surface-affected flows \cite{Ji_Park_Li_Shen_2025}. In the context of pump foiling, it is believed that operation in close proximity to the free surface provides a passive restoring tendency that contributes in maintaining vertical equilibrium. Within a narrow range of submergence depths, hydrofoil lift decreases as the foil approaches the free surface and increases with deeper immersion, a stabilizing mechanism historically exploited by \cite{Alexeev1975} in the design of hydrofoil ships. The simplified description of \cite{rozhdestvensky2023simplified} models the pumped hydrofoil as a forced linear oscillator driven by rider motion and coupled to unsteady lift generation by an oscillating submerged foil. 

Our objective is to attack the pump foil description from the quasi-steady point of view, assuming that the quasi-steady lift and drag properties are given {\it a priori} and that their dependence on the submergence depth fluctuations can be ignored at leading order. We draw inspiration from a series of papers by Ristroph and co-authors \cite{Li_Wang_Ristroph22,Pomerenk_Ristroph_2025} and strive to provide a simplified quasi-steady framework that captures the core physics of self-propulsion acting in pump foil. The simplicity of the model enables us to provide a complete dynamical model: we do not prescribe the kinematics of the foil, but rather use the forces applied by the two feet of the rider as input variables. While we prescribe their sum (and therefore the total applied force) as an open loop control, the torque is prescribed as a proportional control law with the pitch angle as measure. We show that our model is suitable to quantitatively capture pump foil dynamics. 

 Oscillating foil  propulsion is obviously not limited to new recreational sports: from fish swimming to bird flight, propulsion arises through periodic body motions that impart net streamwise momentum to the surrounding fluid, producing a jet-like wake. This thrust-producing wake of an oscillating foil is typically characterized by a reverse Bénard–von Kármán (rBvK) vortex street, distinguished from the classical drag-producing wake by its reversed vortex orientation and associated streamwise momentum jet. Godoy-Diana et al.~\cite{GodoyDiana2008PF,GodoyDiana2009JFM} first documented transitions in the wake of a flapping foil, and  proposed a model for symmetry breaking of the rBvK vortex street, showing that increasing forcing amplitude can induce jet deflection and asymmetric wake structures. Ramananarivo \textit{et al.} \cite{Ramananarivo11, Ramananarivo16} further investigated the role of unsteady kinematic structural timing: in a flexible wing model, efficiency is improved by tuning the phase dynamics of deformation rather than maximizing amplitude via resonance, while in tandem configurations a follower can lock into the leader’s periodic wake, yielding stable spacings and a shared (often increased) cruising speed.

 On one hand, experimental studies demonstrated that appropriately tuned heave-pitch motions can achieve high propulsive efficiencies, comparable to those observed in biological swimmers \cite{Anderson1998}. In such combined motions, pitch lags heave by approximately a quarter period, aligning the instantaneous lift vector to maximize streamwise thrust while minimizing power expenditure. These results established the importance of phase relationships and vortex timing in efficient propulsion.

 On the other hand, a minimal realization of this mechanism is the freely moving harmonically heaving foil \cite{Vandenberghe2004,Vandenberghe2006}, which isolates transverse oscillation as the sole kinematic input. Despite its apparent simplicity, the heaving foil captures the essential interplay between added-mass forcing, unsteady circulation and lift generation and vortex shedding that underlies more complex flapping propulsion. Further analysis of unidirectional flight of a free flapping wing revealed that locomotion may arise through a bifurcation in the coupled fluid-structure system. Thus, symmetry breaking is not merely a secondary wake feature but can constitute the primary mechanism through which propulsion emerges.

A recent scaling analysis by Floryan et al.~\cite{Floryan2017} clarified the dependence of thrust and power coefficients on the reduced frequency and Strouhal number. In heave-dominated regimes, the thrust coefficient scales approximately with the square of the Strouhal number, reflecting the quadratic dependence of added-mass forcing on oscillation amplitude and frequency. These scaling relations unify data across kinematic regimes and provide a predictive structure for propulsive performance. 
Beyond prescribed kinematics, the coupling between body dynamics and fluid forces plays a crucial role in determining the flight modes of falling bodies. Li et al.~\cite{Li_Wang_Ristroph22} examined the influence of the location of the center of mass on the stability and dynamics of thin flat gliders, demonstrating how the mass distribution alters flight modes and force balances. This analysis is very complete in terms of reduced-order modeling and provides a firm ground for our model.

The paper is structured as follows. Section \ref{sec:Problem} defines the system of interest and introduces the modeling assumptions; more detailed component models are provided in subsequent sections. In \S\ref{sec:modeldescription}, we derive the equations of motion from Newton’s second law (force acceleration and moment–angular-acceleration balances) and discuss how the rider's inputs enter the model as control actions. Finally, we present the simulation results and discuss their implications for pump foil dynamics.

\section{Problem formulation \label{sec:Problem}}
We aim to formulate a simple model of the pump foil dynamics with three degrees of freedom, described by the translational coordinates $\mathbf{x}=[x,\ y]^T$ and the rotational coordinate $\theta$ (see Figure~\ref{fig:schematicLiftDrag}), as

\begin{align}
m\ddot{x} = \sum F_x \label{eq:intro1}\\
m\ddot{y} = \sum F_y \label{eq:intro2}\\
I\ddot{\theta} = \sum\tau \label{eq:intro3} 
\end{align}
where  $m$ is the mass, $I$ the moment of inertia. These values can be an effective mass or effective inertia if we consider added mass or added inertia, respectively. The forces $F_x$ and $F_y$ are the forces acting on the reference frame, and $\tau$ is the torque components acting on the pivot point, our chosen origin ($\mathbf{O}$).  
 
\subsection{A note on basic quasi-steady consideration of lift and drag forces on a foil}
\label{sec:basics}

Before detailing the full pump foil model, we briefly summarize the standard quasi-steady description (treating the instantaneous loads as functions of the instantaneous kinematics, i.e., neglecting unsteady effects from time-varying translational and angular velocities) of lift and drag on a foil. This  will be used to model the lift and drag on each of the two foils in the pump foil system providing the force components used in (\ref{eq:intro1})-(\ref{eq:intro3}).

\begin{figure}[!ht]\centering
  \begin{tikzpicture}
      \node[anchor=south west, inner sep=0] (image) at (0,0){
\includegraphics[width=\textwidth, angle=0]{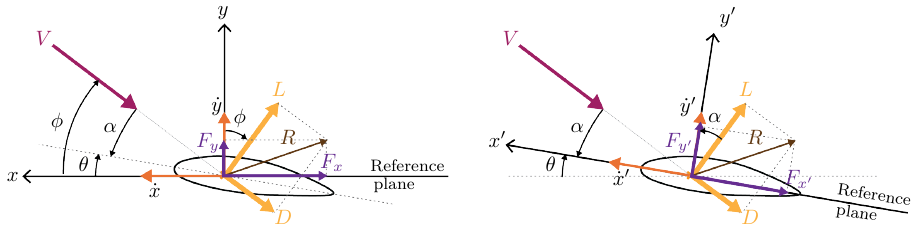}};
    \begin{scope}[x={(image.south east)}, y={(image.north west)}]
      \node[anchor=north west] at (0.0,1.1) {(a)};
      \node[anchor=north west] at (0.5,1.1) {(b)};
    \end{scope}
      \end{tikzpicture}
\caption{Aerodynamic force representation in (a) inertial reference frame ($xy$) and (b) non-inertial frame ($x'y'$): $L$ stands for lift force, $D$ the drag force, $\theta$ the pitch angle, $\phi$ the relative inflow angle (null in $x'y'$ frame), and $\alpha$ the angle of attack (AoA) satisfying the relation $\alpha = \theta-\phi$. It is negative value in this example. The forces $F_x$ and $F_y$ (or $F_{x'}, F_{y'})$ are the projected total forces by lift and drag onto the reference frame $xy$ (or $x'y'$). The incoming velocity $V=\sqrt{\dot{x}^2+\dot{y}^2} = \sqrt{\dot{x}^{'2}+\dot{y}^{'2}}$. } 
\label{fig:schematicLiftDrag}
 \end{figure}

Figure \ref{fig:schematicLiftDrag} sketches the lift ($L$) and drag ($D$) forces acting on a foil \cite{Leishman2016} in two different reference frames. The first is an inertial laboratory frame, and the second is a non-inertial body fixed frame aligned with the chord line of the foil. In both frames, the positive $x$-direction is taken from right to left, the positive $y$-direction is upward, and nose-up rotations are defined as positive $\theta$. 
When the foil moves with horizontal and vertical velocities $\dot{x}$ and $\dot{y}$ (Figure \ref{fig:schematicLiftDrag}a), the effective flow velocity seen by the foil is the vector ${V}$. This velocity makes an angle $\phi$ with the $x$-axis of the inertial $xy$ frame. The angle $\phi$ is referred to as the relative inflow angle or induced angle of attack. The drag force acts along ${V}$, opposing the motion ($\dot{x}$ and $\dot{y}$), while the lift force acts perpendicular to the drag, as illustrated in Figure \ref{fig:schematicLiftDrag}a. The sum of these two forces gives the net hydrodynamic force $R$, which can be decomposed into components $F_x$ and $F_y$ in the $xy$ frame. The same net force can also be projected into the body fixed $x'y'$ frame, yielding components $F_{x'}$ and $F_{y'}$ (Figure \ref{fig:schematicLiftDrag}b).

The angle $\theta$ is the pitch angle, defined as the angle between a reference line on the body (here the chord line of the foil) and the inertial $x$-axis. The angle $\alpha$ is the geometric angle of attack, defined as the angle between the chord line and the incoming flow direction, while $\phi$ is the relative inflow angle introduced above. These angles are related by
\begin{align}
\phi &= \tan^{-1}\!\left(\frac{\dot{y}}{\dot{x}}\right), \label{eq:phi}\\
\alpha &= \theta - \phi. \label{eq:alpha}
\end{align}

In the present analysis, we choose to work in the inertial frame of reference (the “observer-on-the-beach” frame) as shown in Figure \ref{fig:schematicLiftDrag}a, rather than in the non-inertial frame attached to the rider’s foil. The latter can be convenient, in particular for incorporating added mass effects or for certain torque considerations. {We demonstrate in Appendix {\ref{sec:addedMass}} how to include the added mass contribution in the body fixed frame of reference, and show that it has a negligible impact on the results. Therefore, added mass and added inertia are neglected in the main manuscript.}

For a given instantaneous angle of attack $\alpha$, the lift and drag forces acting on the foil, projected onto the $xy$-plane, are
\begin{align}
F_{x} &= -L(\alpha) \sin \phi - D(\alpha) \cos \phi, \label{eq:Fx} \\
F_{y} &=  L(\alpha) \cos \phi - D(\alpha) \sin \phi, \label{eq:Fy}
\end{align}
where
\begin{align}
L(\alpha) &= \frac{1}{2} \rho\, C_L(\alpha)\, A\, V^2, \label{eq:L}\\
D(\alpha) &= \frac{1}{2} \rho\, C_D(\alpha)\, A\, V^2, \label{eq:D}
\end{align}
with $C_L(\alpha)$ and $C_D(\alpha)$ the lift and drag coefficients, respectively, $A$ the planform area (chord length $\times$ span), and $V = \sqrt{\dot{x}^2 + \dot{y}^2}$ the magnitude of the foil velocity in the $xy$-plane. Using
$\sin \phi = \dot{y}/V$ and $\cos \phi = \dot{x}/V$, and omitting the explicit dependence on $\alpha$ for brevity, equations \eqref{eq:Fx}--\eqref{eq:Fy} can be rewritten as
\begin{align}
F_{x} &= -S \big(C_L \dot{y} - C_D \dot{x}\big), \label{eq:Fxs} \\
F_{y} &=  S \big(C_L \dot{x} - C_D \dot{y}\big), \label{eq:Fys} \\
S &= \frac{1}{2} \rho A V.
\end{align}
Equation \eqref{eq:Fxs} shows that, even if there is no forward motion $\dot{x}\simeq 0$, a downward heaving motion with $\dot{y} < 0$ (when the rider pumps the board) can generate positive thrust $F_x$ that enhances the forward velocity (provided that $C_L>0$), as well as a positive net vertical force $F_y$ that can partially or fully balance the rider's weight, as shown in \eqref{eq:Fys}. In the following section, we describe these aerodynamic forces together with the other forces acting in the pump foil model.

It is worth to note that in \citet{Li_Wang_Ristroph22}, the lift coefficients are symmetric with respect to zero angle of attack (AoA), with zero lift at zero AoA. In their convention, the AoA is negative when the wing generates positive lift. The final expression is still correct because the sign is effectively fixed by the velocity terms; however, if the lift coefficient were not symmetric with zero lift at zero AoA (which is usually the case for a good foil), this would lead to an incorrect evaluation of the lift force. Their choice of a negative AoA for positive lift can be understood if one adopts a convention where positive angles correspond to clockwise rotation. In this work, however, we follow the standard aerodynamic convention: angles are positive when the body is nose-up, i.e. $\theta>0$ for nose-up rotation, and $\alpha>0$ for positive-lift (nose-up) configurations.
  
\section{Model description}
\label{sec:modeldescription}
In this section, we formulate a force balance model at the pivot point for the translational and rotational motions of the pump foil. We first describe all relevant translational forces. Whenever these forces do not act through the pivot point, they generate torques, which are then treated in the second part of this section.

\begin{figure}[!ht]
   \centering
  \begin{tikzpicture}
      \node[anchor=south west, inner sep=0] (image) at (0,0){
\includegraphics[width=\textwidth, angle=0]{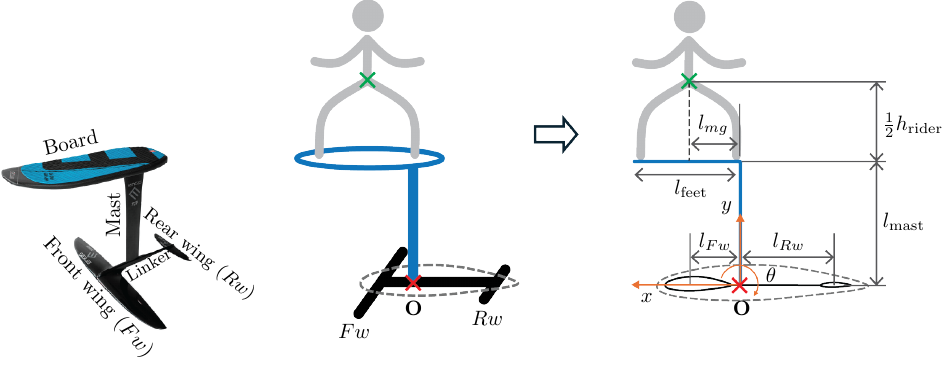}};
    \begin{scope}[x={(image.south east)}, y={(image.north west)}]
          \node[anchor=north west] at (0.0,1.1) {(a)};
      \node[anchor=north west] at (0.3,1.1) {(b)};
    \end{scope}
      \end{tikzpicture}  
 \caption{(a) Typical pump foil shape and the names of components (image source \cite{HipHop98}). (b) Schematic representation of pump foil and rider and relevant dimensions. The pivot point, the rotation center ($\mathbf{O}$), is marked with red cross. The positive directions are chosen as $x$ from right to left, $y$ from bottom to top and $\theta$ the nose up direction. }
       \label{fig:schematic}
 \end{figure}

The translational forces considered are the combined weight of the rider and foil, the pumping force, buoyancy, lift and drag (as described in \S \ref{sec:basics}), and the rotational lift force. Our analysis follows the framework of \citet{Li_Wang_Ristroph22} and the subsequent work of \citet{Pomerenk_Ristroph_2025}. Figure \ref{fig:schematic} shows the model geometry, a typical pump foil configuration, and the notation for the main components.
 
This rider-foil coupled system undergoes the same motion with the exception of an extra term coming from the rider’s relative pumping acceleration:
 \begin{equation}
 \ddot{\mathbf{x}}_{\text {rider }}=\ddot{\mathbf{x}}+\ddot{\mathbf{x}}_{\text {pumping}}.
 \end{equation}
To avoid introducing further unknowns ($\ddot{\mathbf{x}}_{\text {pumping}}$) into the system, we do not model this pumping motion explicitly; instead, we represent its net effect through the forcing term $\mr \ddot{\mathbf{x}}_{\text{pumping}}$ as $\mathbf{F}_{\text{pump}}=[F_{\text{pump},x},\ F_{\text{pump},y}]$:
\begin{equation}
\mf \ddot{\mathbf{x}} + \mr \ddot{\mathbf{x}}_{\text{rider}} = (\mf+\mr)  \ddot{\mathbf{x}} +  \mr \ddot{\mathbf{x}}_{\text{pumping}} = m \ddot{\mathbf{x}} - \mathbf{F}_{\text{pump}},
\end{equation} 
where $m$ is the total mass and $\mathbf{F}_{\text{pump}}$ the extra inertial term due to rider motion relative to $\mathbf{O}$. Note that the rider's muscular action does not appear as a separate force term, since it is internal to the coupled rider-foil system and is represented here only through the prescribed relative pumping motion.  

The main assumptions of the model are as follows:
 \begin{enumerate}[(i)]
   
  \item The hydrodynamic forces are treated as quasi-steady, and only pitching motion about the $z$-axis (normal to the page) is considered. Rolling about the $x$-axis (right-left) and yawing about the $y$-axis (bottom-up) are neglected.
  
  \item The hydrodynamic (lift and drag) forces are generated only by the front wing ($Fw$) and rear wing ($Rw$). Contributions from the mast and the linker between the mast and wings are neglected.
  
  \item  Viscous (skin-friction) drag on the foil, mast, and wing linker is neglected; only pressure-induced lift and drag are retained.
  
  \item For each wing, the centers of pressure, buoyancy, and volume are assumed to be nearly coincident compared with the total length. Thus, the distance from the point $\mathbf{O}$ is represented by a single lever arm for each wing, $l_{Fw}$ and $l_{Rw}$.
  
  \item All forces are assumed to act at a single effective point on each body.
\end{enumerate}

\subsection{Translation forces, $\sum F_x$, $\sum F_y$}
\subsubsection{Buoyancy and rider's contribution}
\begin{figure}[ht]\centering
  \begin{tikzpicture}
    \node[anchor=south west, inner sep=0] (image) at (0,0){
      \includegraphics[width=0.4\textwidth, angle=0]{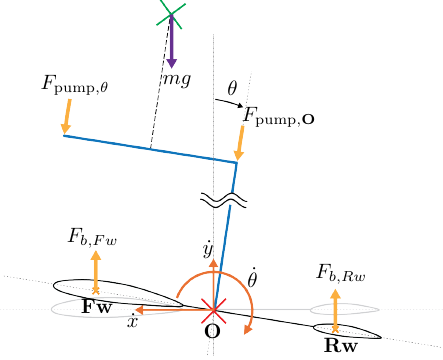}};
    \begin{scope}[x={(image.south east)}, y={(image.north west)}]
    \end{scope}
  \end{tikzpicture}
  \caption{Forces due to the weight and pumping, and due to the buoyancy of the front and rear wings.}
  \label{fig:mass_buoyancy}
\end{figure}

The first and most obvious contributions are the weight and the hydrostatic buoyancy acting on the foil, as shown in Figure \ref{fig:mass_buoyancy}. Since the rider mass dominates the total mass, the system center of mass is approximated by the rider’s body center, and the weight is therefore taken to act at that point.

We assume that the rider exerts a pumping force perpendicular to the board. The resulting vertical and horizontal force components are then
\begin{align}
F_{\text{mass},y} &= - m g, \\
F_{b,y} &= F_{b,Fw} + F_{b,Rw} 
       = (m_{b,Fw} + m_{b,Rw})\, g, \\
F_{\text{pump},x} &= F_{\text{pump},tot} \sin\theta, \\
F_{\text{pump},y} &= F_{\text{pump},tot} \cos\theta,
\end{align}
where $F_{\text{pump},tot} = F_{\text{pump},\mathbf{O}} + F_{\text{pump},\theta}$ is the total pumping force, decomposed into the component applied at the pivot point $\mathbf{O}$ and the component associated with pitching. The rider’s contribution is discussed in detail in a dedicated section.

\subsubsection{Lift and Drag based forces}
There are two lift- and drag-based force contributions acting on the system: those generated by the front wing ($Fw$) and by the rear wing ($Rw$). The analysis follows the same exercise demonstrated previous section on lift and drag, but here the relevant local velocities at the centers of pressure of each wing must be used.

\begin{figure}[!ht]\centering
   \begin{tikzpicture}
      \node[anchor=south west, inner sep=0] (image) at (0,0){
\includegraphics[width=\textwidth, angle=0]{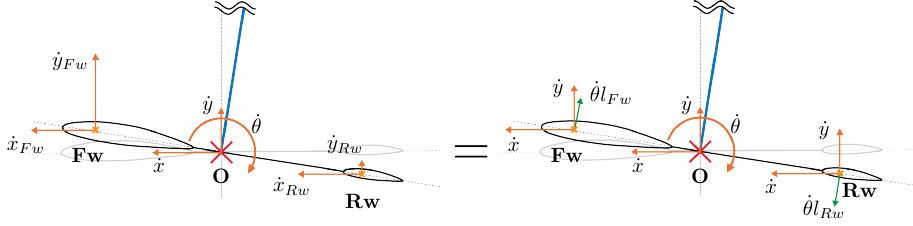}};
    \begin{scope}[x={(image.south east)}, y={(image.north west)}]

    \end{scope}
      \end{tikzpicture}
  \caption{Velocities at the front and rear wings. The local velocities at each point are modified by the rotational velocity induced by the pitching motion.}
      \label{fig:lift_drag_velocity}
 \end{figure}

Since the lift and drag forces are defined based on the velocity at the pressure center of each wing, we first correct the velocity from the pivot point $\mathbf{O}$ to the corresponding local positions. This is illustrated in Figure \ref{fig:lift_drag_velocity}.  
 Let the pressure center of the front wing be located at point $\mathbf{Fw}$. The local velocity there is $
  [\dot{x}_{Fw}, \dot{y}_{Fw}],$
with magnitude
$  V_{Fw} = \sqrt{\dot{x}_{Fw}^2 + \dot{y}_{Fw}^2}.$
The lift and drag forces on the front wing must be evaluated using these local velocities. Similarly, the pressure center of the rear wing is at point $\mathbf{Rw}$ with local velocity $[\dot{x}_{Rw}, \dot{y}_{Rw}]$ and magnitude $V_{Rw}$.

The local velocities differ from those at the pivot point $\mathbf{O}$ because of the pitching motion. The rotational contribution at a distance $l$ from $\mathbf{O}$ is of order $l \dot{\theta}$. In the $xy$ coordinate system, the total velocities at the front and rear wings can be written  as function of $\dot{x}$ and $\dot{y}$:
\begin{align}
\dot{x}_{Fw} &= \dot{x} -l_{Fw} \sin \theta,\quad \dot{y}_{Fw} = \dot{y} +l_{Fw} \cos \theta, \label{eq:velFw}\\
\dot{x}_{Rw} &= \dot{x} +l_{Rw} \sin \theta,\quad \dot{y}_{Rw} = \dot{y} -l_{Rw} \cos \theta.  \label{eq:velRw}
\end{align} 

The angle relations defined in (\ref{eq:phi})--(\ref{eq:alpha}) hold with the global velocities replaced by the local ones. For example, for the front wing
\begin{equation*}
  \phi_{Fw} = \tan^{-1}\!\left(\frac{\dot{y}_{Fw}}{\dot{x}_{Fw}}\right),
  \quad
  \alpha_{Fw} = \theta - \phi_{Fw},
\end{equation*}
and similarly for the rear wing.

\begin{figure}[!ht]
 \centering\includegraphics[width=0.8\textwidth, angle=0]{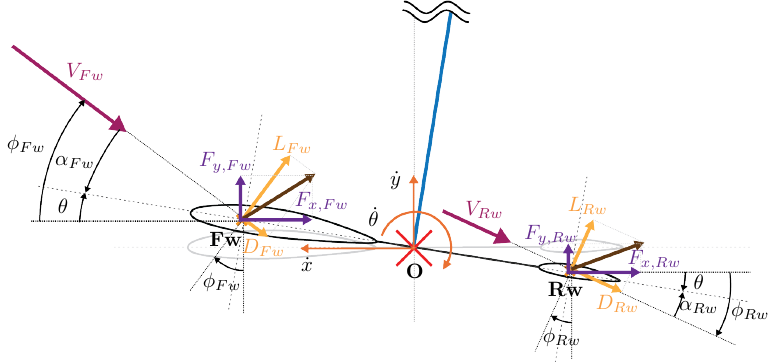}
  \caption{Lift ($L$) and drag ($D$) forces acting at $\mathbf{Fw}$ and $\mathbf{Rw}$ and their projections $F_{x,\cdot}$ and $F_{y,\cdot}$ in the $xy$ frame.}
      \label{fig:lift_drag_forces}
 \end{figure}

Figure \ref{fig:lift_drag_forces} shows the lift and drag forces acting on each wing and their projections onto the $xy$ components.

Following the same reasoning as in (\ref{eq:L})--(\ref{eq:Fys}), the front-wing force components in the $xy$ frame are
\begin{align}
 F_{x,Fw} &=  -S_{Fw}(C_{L,Fw} \dot{y}_{Fw} - C_{D,Fw} \dot{x}_{Fw}), \\
 F_{y,Fw} &=   S_{Fw}(C_{L,Fw} \dot{x}_{Fw} - C_{D,Fw} \dot{y}_{Fw}), \\
 S_{Fw} =& \frac{1}{2} \rho A_{Fw} V_{Fw}. 
   \end{align}
The same expressions apply to the rear wing:
\begin{align}
 F_{x,Rw} &=  -S_{Rw}(C_{L,Rw} \dot{y}_{Rw} - C_{D,Rw} \dot{x}_{Rw}), \\
 F_{y,Rw} &=  S_{Rw}(C_{L,Rw} \dot{x}_{Rw} - C_{D,Rw} \dot{y}_{Rw}), \\
 S_{Rw} =& \frac{1}{2} \rho A_{Rw} V_{Rw}. 
 \end{align}

\subsubsection{Lift from rotation}
\label{sec:rotational-lift-force}
The last translational force contribution is the lift generated by rotation of the foil about the axis at $\mathbf{O}$, a Magnus-type effect. According to the Kutta-Joukowski theorem, when a body rotates and thereby generates a circulation $\Gamma$ about the $z$-axis, it experiences a lift force  given, in the present notation, by 
\begin{align}
F_{x,RL} &= - \rho\,\Gamma\,\dot{y}, \\
F_{y,RL} &= \phantom{-}\rho\,\Gamma\,\dot{x}.
\end{align}

\begin{figure}[ht]\centering
  \begin{tikzpicture}
    \node[anchor=south west, inner sep=0] (image) at (0,0){
      \includegraphics[width=0.4\textwidth, angle=0]{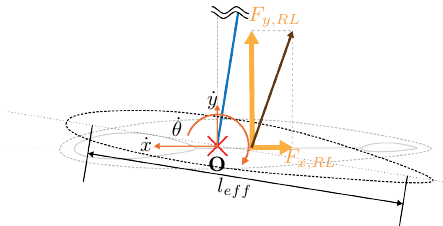}};
  \end{tikzpicture}
  \caption{Forces associated with rotational (Magnus-type) lift.}
  \label{fig:lift_rotational}
\end{figure}

For a rotating cylinder, the circulation is simply $\Gamma = - 2 \pi r V_{\theta}$, where $r$ is the radius and $V_{\theta}$ the circumferential velocity. In a more general form suitable for our configuration. \citet{Pomerenk_Ristroph_2025} wrote the expression for a thin plate
\[
  \Gamma = - C_R\,l_{\text{eff}}\left(\frac{l_{\text{eff}}}{2} \dot{\theta}\right),
\]
where $C_R$ is a rotational-lift coefficient, $l_{\text{eff}}$ is an effective length scale, as sketched in Figure \ref{fig:lift_rotational}. This leads to the following expression for the rotational lift contribution in the $xy$ frame:
\begin{align}
F_{x,RL} &= - S_{RL}\,\dot{\theta}\,\dot{y}, \label{eq:FxRL} \\
F_{y,RL} &= \phantom{-}S_{RL}\,\dot{\theta}\,\dot{x}, \\
S_{RL}   &= \rho\,C_R\,A_{RL}\,l_{\text{eff}},
\label{eq:FyRL}
\end{align}
where $A_{RL}$ is an effective planform area associated with the rotational lift. As seen in Fig {\ref{fig:lift_rotational}}, the value of $A_{RL}$ is typically of order $A_{Fw}+A_{Rw}$.

Because $l_{\text{eff}}$ and $A_{RL}$ are not known, we can distribute the rotational lift contribution over the two wings using the known lever arms $l_{Fw}$ and $l_{Rw}$ and the local velocities defined in (\ref{eq:velFw})–(\ref{eq:velRw}):
\begin{align}
F_{x,RL} &= -\left(S_{RL,Fw}\,\dot{y}_{Fw} + S_{RL,Rw}\,\dot{y}_{Rw}\right)\dot{\theta}, \\
F_{y,RL} &= \phantom{-}\left(S_{RL,Fw}\,\dot{x}_{Fw} + S_{RL,Rw}\,\dot{x}_{Rw}\right)\dot{\theta}, \\
S_{RL,Fw} &= \tfrac{1}{2}\rho\,C_R\,A_{Fw}\,l_{Fw}, 
\qquad
S_{RL,Rw} = \tfrac{1}{2}\rho\,C_R\,A_{Rw}\,l_{Rw},
\end{align}
where $\dot{x}_{Fw}, \dot{y}_{Fw}$ and $\dot{x}_{Rw}, \dot{y}_{Rw}$ are the corrected local velocities at the front and rear wings, respectively. Using (\ref{eq:velFw})–(\ref{eq:velRw}), this reduces to
\begin{align}
F_{x,RL} &= F_{x,RL,Fw} + F_{x,RL,Rw}
          = - S_{RL,Fw}\,\dot{\theta}\,(\dot{y} + l_{Fw}\cos\theta)
            - S_{RL,Rw}\,\dot{\theta}\,(\dot{y} - l_{Rw}\cos\theta),\label{eq:rotLift_indx}\\
F_{y,RL} &= F_{y,RL,Fw} + F_{y,RL,Rw}
          = \phantom{-}S_{RL,Fw}\,\dot{\theta}\,(\dot{x} - l_{Fw}\sin\theta)
            + S_{RL,Rw}\,\dot{\theta}\,(\dot{x} + l_{Rw}\sin\theta).
\label{eq:rotLift_indy}
\end{align}

\subsection{Torque balance}

In this section, we evaluate the torques acting about the pivot point $\mathbf{O}$. All the linear forces discussed in previous section will exert torque if they are not aligned on the radial direction of $\mathbf{O}$. That are buoyancy, lift-drag forces, rotational lift, pure rotation driven and rider's contribution.  


\subsubsection{Torque due to buoyancy}
The torque generated by buoyancy is given by the tangential components of the front- and rear-wing buoyancy forces (see Figure \ref{fig:mass_buoyancy}). With the convention that positive torque corresponds to nose-up rotation, we obtain
\begin{equation}
  \tau_{B} = \bigl(F_{b,Fw}\,l_{Fw} - F_{b,Rw}\,l_{Rw}\bigr)\cos\theta.
\end{equation}

\subsubsection{Torques from translational lift and drag}
\begin{figure}[ht]\centering
  \begin{tikzpicture}
    \node[anchor=south west, inner sep=0] (image) at (0,0){
      \includegraphics[width=0.8\textwidth, angle=0]{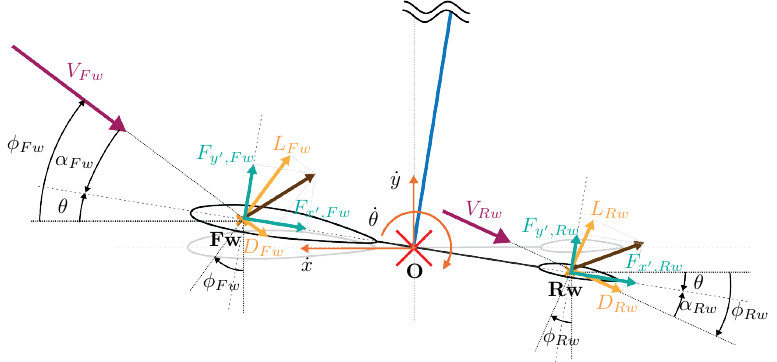}};
  \end{tikzpicture}
  \caption{Translational torque due to lift and drag forces. Only the tangential components $F_{y'}$ (blue-green arrows) contribute to the torque about the pivot point $\mathbf{O}$.}
  \label{fig:torque_T}
\end{figure}

To evaluate the torques due to lift and drag, we project the hydrodynamic forces onto the directions tangential and normal to the circular motion about $\mathbf{O}$. This is equivalent to expressing the forces in the wing linker frame of reference $x'y'$ introduced in Figure \ref{fig:schematicLiftDrag}. The construction is illustrated in Figure \ref{fig:torque_T}. Note that the only difference from Figure \ref{fig:lift_drag_forces}, is the projection of forces on the frame of reference $x'y'$.

The transformation from the inertial $xy$ components to the board frame $x'y'$ for the front wing reads
\begin{align}
F_{x',Fw} &= F_{x,Fw}\cos\theta + F_{y,Fw}\sin\theta, \\
F_{y',Fw} &= -F_{x,Fw}\sin\theta + F_{y,Fw}\cos\theta,
\end{align}
and the same transformation applies to the rear wing, giving $F_{x',Rw}$ and $F_{y',Rw}$. In this frame, the $y'$-components are tangential to the circular motion about $\mathbf{O}$ and therefore generate torque, whereas the $x'$-components are radial and do not.

The resulting torque $\tau_T$ from the front and rear wings is thus
\begin{equation}
  \tau_T = \tau_{T,Fw} + \tau_{T,Rw}
         = F_{y',Fw}\,l_{Fw} - F_{y',Rw}\,l_{Rw}.
\end{equation}

For small pitch angles, $\theta \ll 1$, the board-fixed frame is nearly aligned with the inertial frame, so that $F_{y',\cdot} \simeq F_{y,\cdot}$. In this limit, the torque simplifies to
\begin{equation}
  \tau_T \simeq F_{y,Fw}\,l_{Fw} - F_{y,Rw}\,l_{Rw}.
\end{equation}

\subsubsection{Torque from pure rotation}
\begin{figure}[ht]\centering
  \begin{tikzpicture}
    \node[anchor=south west, inner sep=0] (image) at (0,0){
      \includegraphics[width=0.6\textwidth, angle=0]{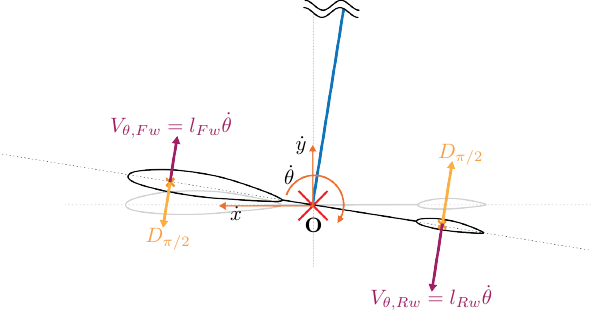}};
  \end{tikzpicture}
  \caption{Torque due to purely rotational motion, arising from the drag force $D_{\pi/2}$ at $90^\circ$ angle of attack.}
  \label{fig:torque_Rot}
\end{figure}

When the body undergoes pure pitching motion, the front and rear wings experience a resistive torque associated with drag acting at an effective angle of attack of $\pi/2$, as illustrated in Figure \ref{fig:torque_Rot}. At the front wing, the local circumferential speed is
\[
  V_{\theta,Fw} = l_{Fw}\,\dot{\theta},
\]
with an analogous expression at the rear wing.

The net rotational drag torque is then
\begin{equation}
  \tau_R = -D_{\pi/2,Fw}\,l_{Fw} - D_{\pi/2,Rw}\,l_{Rw},
\end{equation}
where
\begin{equation}
  D_{\pi/2,Fw} 
  = \tfrac{1}{2}\,C_{D,\pi/2}\,\rho\,A_{Fw}\,V_{\theta,Fw}^2
  = \tfrac{1}{2}\,C_{D,\pi/2}\,\rho\,A_{Fw}\,l_{Fw}^2\,\dot{\theta}^2,
\end{equation}
and similarly for the rear wing. Here $C_{D,\pi/2}$ is the drag coefficient at $\alpha = \pi/2$.

Because drag always opposes the motion, the torque must change sign with $\dot{\theta}$. This dependence can be compactly written using the factor $\dot{\theta}\,|\dot{\theta}|$, which ensures that the torque is always directed opposite to the angular velocity (with the negative sign in front). Combining the contributions from both wings, we obtain
\begin{equation}
  \tau_R
  = -\tfrac{1}{2}\,\rho\,C_{D,\pi/2}\,
    \bigl(A_{Fw}\,l_{Fw}^3 + A_{Rw}\,l_{Rw}^3\bigr)\,
    \dot{\theta}\,|\dot{\theta}|.
\end{equation}
A similar nonlinear rotational drag term appears in \cite{Anderson_Wang_2005,Li_Wang_Ristroph22,Pomerenk_Ristroph_2025}.

\subsubsection{Torque from rotational lift}
\begin{figure}[ht]\centering
  \begin{tikzpicture}
    \node[anchor=south west, inner sep=0] (image) at (0,0){
      \includegraphics[width=0.45\textwidth, angle=0]{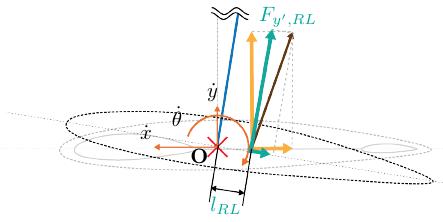}};
  \end{tikzpicture}
  \caption{Torque generated by rotational lift acting at a distance $l_{RL}$ from the pivot point $\mathbf{O}$.}
  \label{fig:torque_Rotlift}
\end{figure}

In \citet{Pomerenk_Ristroph_2025}, it is argued that the resultant rotational lift force need not act exactly through the pivot point $\mathbf{O}$, but may instead be applied at some effective distance from it $l_{RL}$, as sketched in Figure \ref{fig:torque_Rotlift}. Although the precise location of this center of action is uncertain, it is reasonable to allow for a generic offset in order to keep the formulation as general as possible.

If the net rotational-lift force $F_{y',RL}$ acts at a distance $l_{RL}$ along the board-fixed frame, its contribution to the torque about $\mathbf{O}$ is
\begin{equation}
  \tau_{RL} = l_{RL}\,F_{y',RL}
            = l_{RL}\bigl(-F_{x,RL}\sin\theta + F_{y,RL}\cos\theta\bigr),
\end{equation}
where $(F_{x,RL},F_{y,RL})$ are the components in the inertial $xy$ frame and $F_{y',RL}$ is the tangential component in the board-fixed frame. 

In our case, we have no reliable way to estimate $l_{RL}$, but it is possible to  
distribute the rotational-lift contribution over the two wings, using the individual moment arms $l_{Fw}$ and $l_{Rw}$ as in (\ref{eq:rotLift_indy}):
\begin{equation}
  \tau_{RL} = l_{Fw} F_{y',RL,Fw} - l_{Rw} F_{y',RL,Rw},
\end{equation}
where $F_{y,RL,Fw}$ and $F_{y,RL,Rw}$ are the rotational-lift contributions from the front and rear wings defined in (\ref{eq:rotLift_indx})--(\ref{eq:rotLift_indy}).

\subsection{Rider's control}
\label{sec:ridercontrol}
In this section, we consider the rider's internal actuation relative to the coupled rider-foil system. The actuation is represented by an effective pumping load associated with the rider's relative motion:
\begin{align} 
F_{\text{pump},x} &= F_{\text{pump},tot}\sin \theta, \\
F_{\text{pump},y} &= F_{\text{pump},tot}\cos \theta. 
\end{align}
This load is further decomposed into front- and rear-foot contributions $F_{\text{pump},tot} = F_{\text{pump},\mathbf{O}} + F_{\text{pump},\theta}$ as shown in Figure \ref{fig:mass_buoyancy}. Their sum drives the translational pumping, while the pitching torque $
\tau_{\text{rider}}$ is assumed to be produced primarily by the front foot force as the rear foot is assumed to be on the pivot point $\mathbf{O}$ (and therefore produces no moment about $\mathbf{O}$) and the rider's center of gravity is assumed to be aligned vertically above $\mathbf{O}$ at any time, as explained in more details in section \ref{sec:ridertorquecontrol}.

It is worth noting that, while we prescribe the total force between the two feet, their respective share remains undetermined so far. We will assume it to be closed-loop controlled, as discussed next.

\subsubsection{Rider's pumping force}

For the translational dynamics, the only relevant rider forcing is the total pumping force $F_{\text{pump},tot}(t)$. We model it as a sinusoidal function,
\begin{equation}
  F_{\text{pump},tot}(t) = A_{\text{pump},tot}\,\sin(\omega t),
\end{equation}
where $\omega$ is the angular pumping frequency in $\mathrm{rad/s}$. 

\begin{figure}[ht]
  \centering
  \includegraphics[width=0.8\textwidth]{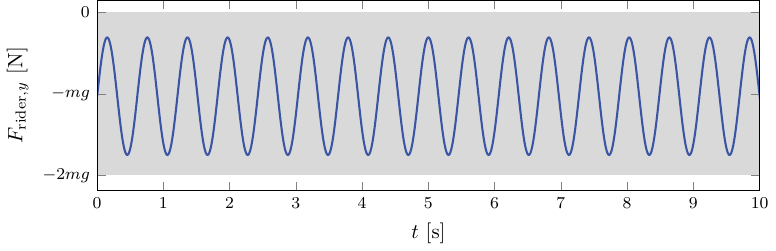}
  \caption{Example pump force and its combination with the rider's weight $F_{\text{pump},y}-mg$, with $\theta=0$. The gray region indicates the admissible range of pumping amplitudes.}
  \label{fig:forcing}
\end{figure}

For the pumping amplitude $A_{\text{pump},tot}$, we assume that the vertical motion is comparable to a jumping motion. In vertical jumps, humans can transiently exert vertical forces of order a few times their own body weight \citep{Bobbert1991}. Motivated by this, we restrict the pumping amplitude to be of order $m g$, i.e. $|A_{\text{pump},tot}| \lesssim m g$, and tune $A_{\text{pump},tot}$ within this range so that the rider maintains stably above the water. An example of the total rider's forcing is shown in Figure \ref{fig:forcing} for $A_{\text{pump},tot} = 0.7\,m g$ with pumping frequency of $1.65$ Hz ($\omega =  2 \pi \times 1.65\ \mathrm{rad/s}$).

The total pumping force can be further decomposed into a component applied at the pivot point $\mathbf{O}$ (rear foot) and a component associated with pitch control (front foot):
\begin{align}
  F_{\text{pump},tot}(t)
  &= F_{\text{pump},\theta}(t) + F_{\text{pump},\mathbf{O}}(t) = A_{\text{pump},tot}\,\sin(\omega t).
   \label{eq:totalforce}
\end{align}
We allow both forces $F_{\text{pump},\theta}(\theta)$ and $F_{\text{pump},\mathbf{O}}(\theta)$ to vary with the pitch angle $\theta$, but their sum is constrained to  the prescribed total amplitude,
This choice keeps the axial pumping amplitude fixed (we tune only $A_{\text{pump},tot}$), while allowing the torque-generating component $F_{\text{pump},\theta}(\theta)$ to be adjusted as a function of $\theta$.

\subsubsection{Rider's torque control} 
\label{sec:ridertorquecontrol}
In this section, we specify the torque applied by the rider. Quantifying this control input is particularly challenging, because in practice it is intimately linked to balance and to the rider's continuous stabilization of the system. Since the authors are not experienced pump foil riders, the assumptions introduced below should be regarded only as a first modeling attempt and should be challenged and refined through future measurement campaigns.    
  Our working assumptions are:

\begin{enumerate}[(i)]
  \item The motion of the rider's arms is not modeled explicitly \footnote{In practice, the rider typically swings the arms fore-aft about the torso with approximately half the frequency of the vertical pumping motion. This motion is important for stable riding (if the rider were not allowed to move the arms, it would be very difficult to ride), but we assume that it mainly contributes to stabilizing other rotational degrees of freedom, such as roll about the $x$-axis, and has a secondary effect on pitch about $y$.}.
  
  \item The rider’s control is prescribed, as if performed by a programmed robot (proportional type control).
  
  \item The sinusoidal forces $F_{\text{pump},\theta}$ and $F_{\text{pump},\mathbf{O}}$ are perfectly synchronized: there is no phase or timing difference between the pumping motions of the two legs.
\end{enumerate}

\begin{figure}[ht]\centering
   \begin{tikzpicture}
      \node[anchor=south west, inner sep=0] (image) at (0,0){
\includegraphics[width=0.7\textwidth, angle=0]{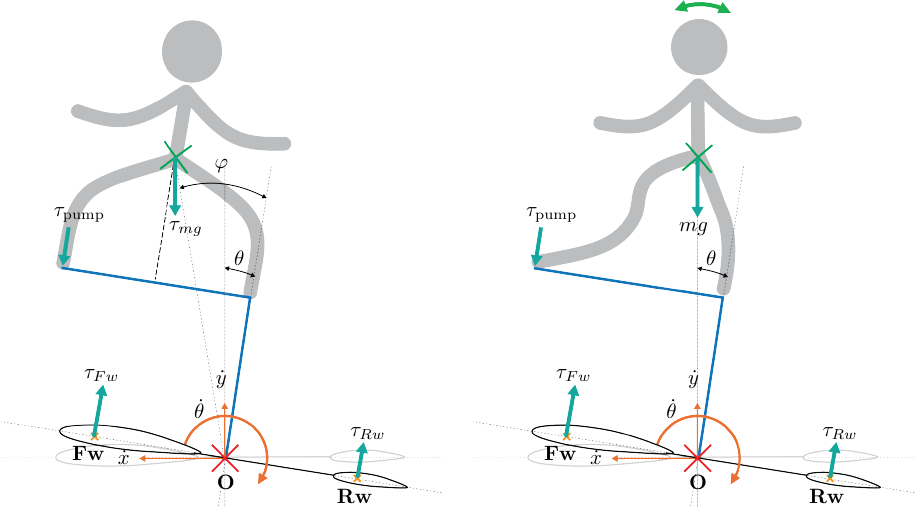}};
    \begin{scope}[x={(image.south east)}, y={(image.north west)}]
      \node[anchor=north west] at (0.0,1.0) {(a)};
      \node[anchor=north west] at (0.5,1.0) {(b)};
    \end{scope}
      \end{tikzpicture}
  \caption{Assumed rider mass-control mechanism. (a) The rider does not adjust his/her center of mass with respect to pitch variations, resulting in a finite torque $\tau_{mg}$. (b) The rider actively repositions the center of mass so that it lies on the rotation axis, leading to $\tau_{mg} \approx 0$.}
  \label{fig:Control_rider2}
\end{figure}

In the present study, we only consider the torque control by the front foot of the rider. This means that we neglect the torque from the rider's mass (see Figure \ref{fig:Control_rider2}a) assuming a situation closer to that depicted in Figure \ref{fig:Control_rider2}b: the rider  adjusts his posture so that the center of mass remains approximately aligned with the rotation axis. In this idealized limit, the weight acts through $\mathbf{O}$ and the associated torque vanishes $ \tau_{mg} = 0$.

As discussed in the rider's pumping-force section and illustrated in Figure \ref{fig:mass_buoyancy}a, only the component $F_{\text{pump},\theta}$ contributes to the torque about $\mathbf{O}$. This torque is simply
\[
  \tau_{\text{rider}} = F_{\text{pump},\theta}\,l_{\text{feet}},
\]
where $l_{\text{feet}}$ is the horizontal distance between the pivot point $\mathbf{O}$ and the rider’s front foot or the distance between front and rear feet (the effective lever arm).

More precisely, we prescribe the pumping force contributing to torque generation as
\begin{equation}
 F_{\text{pump},\theta}= A_{\text{pump},\theta}(\theta)\,\bigl(\sin(\omega t) - 1\bigr),
  \label{eq:taupump}
\end{equation}
 while the force acting on the rear foot $F_{\text{pump},\mathbf{O}}$ is given by \eqref{eq:totalforce}. The factor $\sin(\omega t) - 1$ reflects the assumption that the rider can only exert negative torque (pushing down with the front foot), but cannot pull to generate a positive force; the maximum force, and therefore the maximum torque, is thus zero.
\begin{figure}[!ht]
  \centering
  \includegraphics[width=0.7\textwidth]{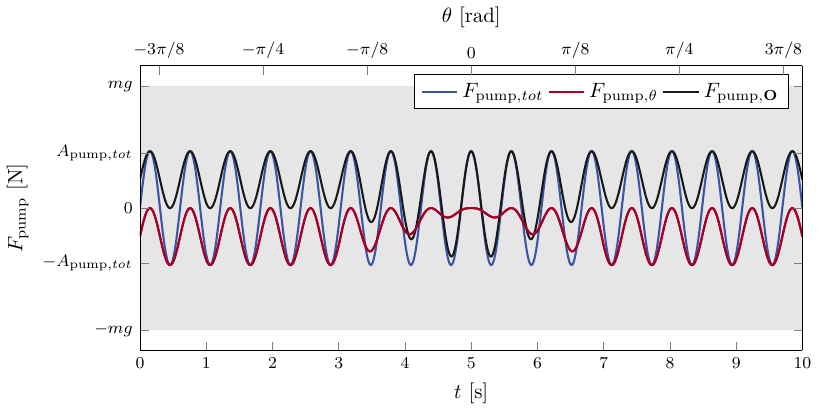}
  \caption{Example amplitudes of pumping forces $F_{\text{pump},\theta}$, $F_{\text{pump},\mathbf{O}}$, and their sum $F_{\text{pump},{tot}}$, for time-varying $\theta(t)$. In this example, $k_\theta = A_{\text{pump},{tot}}/(1\ \mathrm{rad})$ is used. Note that, for any sign of $\theta$, $F_{\text{pump},\theta}$ only takes negative values in our convention, and therefore always generates a torque opposing the nose-up rotation.}
  \label{fig:torque}
\end{figure}
To specify the control amplitude $A_{\text{pump},\theta}(\theta)$, we assume that it increases with the magnitude of the pitch angle, up to the limit set by the total available pumping amplitude of one foot:
\begin{equation}
A_{\mathrm{pump},\theta} = \min\bigl(k_{\theta}\,|\theta|,\; \tfrac{1}{2} A_{\mathrm{pump},tot}\bigr),
\label{eq:taupumpamp}
\end{equation}
where $k_\theta$ is a proportionality constant (with units of force per radian) that sets how sensitively the rider responds to pitch deviations. 

Figure \ref{fig:torque} illustrates the amplitudes of the individual pumping forces $F_{\text{pump},\theta}$ and $F_{\text{pump},\mathbf{O}}$, and their sum $F_{\text{pump},{tot}}$ with an example with time varying $\theta(t)$. In this example, we use $k_\theta = A_{\text{pump},{tot}}/(1\ \mathrm{rad})$. Note that, regardless of the sign of $\theta$, the force $F_{\text{pump},\theta}$ is always negative in our sign convention and thus always produces a nose-down torque. In addition, the sum of the two pitch dependent forces $F_{\text{pump},\theta}(\theta)$ and $F_{\text{pump},\mathbf{O}}(\theta)$ are always bounded as $F_{\text{pump},{tot}} = A_{\text{pump}, tot}\sin (\omega t)$.
This means that the rider's pitch control is modeled solely through the redistribution of the pumping force between $\mathbf{O}$ and the front leg.

\subsection{Final expression}
We gather all the expressions here: 
 \begin{align}
m\ddot{x} &=  F_{x,Fw} + F_{x,Rw} + F_{x,RL} + F_{\text{pump},x}, \label{eq:finalx}\\
m\ddot{y} &=  F_{b,y} -mg + F_{y,fw} + F_{y,Rw} + F_{y,RL} + F_{\text{pump},y},\label{eq:finaly}\\
I\ddot{\theta} &=\tau_{\text{rider}} + \tau_B + \tau_T + \tau_R +\tau_{RL},\label{eq:finalI}
\end{align}
where 
\begin{align}
&\begin{bmatrix}
  F_{\text{pump},x}\\  F_{\text{pump},y} \end{bmatrix} = A_{\text{pump},tot}\sin(\omega t) \begin{bmatrix}
  \sin \theta\\ \cos \theta \end{bmatrix} \simeq A_{\text{pump},tot}\sin(\omega t) \begin{bmatrix}
  0\\   1
\end{bmatrix}, \\
& F_{b,y} =(m_{\text {b},Fw} +m_{\text {b},Fw}) g, \\
& F_{\mathrm{mass},y} = -mg, \\
&\begin{bmatrix}
  F_{x,Fw}\\   F_{y,Fw} \end{bmatrix} = S_{Fw} \begin{bmatrix}
 C_{D,Fw}  & -C_{L,Fw}  \\
 C_{L,Fw}  & -C_{D,Fw} 
\end{bmatrix}
\begin{bmatrix}
\dot{x}\\[2pt]
\dot{y}
\end{bmatrix} - S_{Fw}l_{Fw}
\begin{bmatrix}
C_{D,Fw} & C_{L,Fw}  \\[2pt]
C_{L,Fw} & C_{D,Fw}  
\end{bmatrix} \begin{bmatrix}
  \sin \theta \\ \cos \theta \end{bmatrix} , \\
&\begin{bmatrix}
  F_{x,Rw}\\  F_{y,Rw} \end{bmatrix} =S_{Rw} \begin{bmatrix}
  C_{D,Rw} & -C_{L,Rw} \\
  C_{L,Rw} & -C_{D,Rw}
\end{bmatrix}
\begin{bmatrix}
\dot{x}\\[2pt]
\dot{y}
\end{bmatrix} +
S_{Rw}l_{Rw} \begin{bmatrix}
 C_{D,Rw} & C_{L,Rw}  \\[2pt]
 C_{L,Rw} & C_{D,Rw} \end{bmatrix} \begin{bmatrix}
  \sin \theta\\ \cos \theta \end{bmatrix} , \\
&\begin{bmatrix}
  F_{x,RL}\\  F_{y,RL} \end{bmatrix} = (S_{R L, F w}+S_{R L, R w}) \dot{\theta} \begin{bmatrix}
  -\dot{y} \\ \dot{x} \end{bmatrix} + (S_{R L, R w} l_{R w}-S_{R L, F w} l_{F w}) \dot{\theta} \begin{bmatrix}
  -\cos \theta \\ \sin \theta \end{bmatrix}.
\end{align}

Then, the torque terms 
\begin{align}
\tau_{\text{rider}} &= A_{\text{pump},\theta}(\theta) (\sin(\omega t)-1)l_{\text{feet}},\\
\tau_B &=g (m_{b,Fw}l_{Fw} - m_{b,Rw}l_{Rw})\cos \theta,\\
\tau_T &= (-F_{x,Fw} \sin \theta + F_{y,Fw} \cos \theta) l_{Fw}- (F_{x,Rw} \sin \theta - F_{y,Rw} \cos \theta)l_{Rw},\\
\tau_R &= -\frac{1}{2}\rho C_{D,\pi/2} (A_{Fw} l_{Fw}^3+ A_{Rw} l_{Rw}^3) \dot{\theta} | \dot{\theta}|,\\
\tau_{RL}  &=  l_{Fw} F_{y',RL, Fw} - l_{Rw}  F_{y',RL, Rw}.
\end{align}

Note that the only control parameters which should be tuned are the amplitudes of pumping forces $A_{\text{pump},tot}$ and $A_{\text{pump},\theta}(\theta)$ (or $k_{\theta}$). If we restrict the rider to exert either the full (negative) pumping force or none at all on the front foot as in the example of Figure \ref{fig:torque}, now the only remaining control parameter is $A_{\text{pump},tot}$.

\section{Results}
In this section, we present the solutions of the dynamical system given by (\ref{eq:finalx})--(\ref{eq:finalI}). The physical parameters are taken from \cite{PIS2025} and from the pump foil manufacturers' specifications \cite{axis_complete_foils}; their values are summarized in Table \ref{tab:params}.

To simplify further, we also use the measured pumping frequency. Typical pumping frequencies are in the range $f \in 1$ to $3 \mathrm{Hz}$ \cite{PIS2025,rozhdestvensky2023simplified}. 
 From a student semester project conducted in 2025 spring at Lake Leman \cite{PIS2025}, we indicated typical pumping frequencies around $1.65~\mathrm{Hz}$ for a rider of mass $\sim 70~\mathrm{kg}$ and this value is used in the following. The dependency on the pumping frequency is studied in Appendix \ref{sec:depfreq}.

\begin{table}[!ht]\centering
  \caption{Model parameters used in the simulations.}
  \label{tab:params}
  \begin{tabular}{llll}
Description  & Symbol & Value & Unit \\ \hline
Foil mass   & $m_\mathrm{foil}$      & $2$                 & kg \\
Total mass   & $m$                    & $72$                & kg \\
Gravity   & $g$                    & $9.81$              & m\,s$^{-2}$ \\
Density   & $\rho$                 & $1000$              & kg\,m$^{-3}$ \\
Chord, $Fw$ & $c_{Fw}$        & $0.132$             & m \\
Span, $Fw$ & $b_{Fw}$        & $1.05$              & m \\
Area, $Fw$ & $A_{Fw}=b_{Fw}c_{Fw}$ & $0.139$ & m$^{2}$ \\
Volume, $Fw$ & $\mathcal{V}_{Fw}$        & $1.2\times10^{-3}$  & m$^{3}$ \\
Chord, $Rw$ & $c_{Rw}$        & $0.06$              & m \\
Span, $Rw$ & $b_{Rw}$        & $0.50$              & m \\
Area, $Rw$ & $A_{Rw}=b_{Rw}c_{Rw}$ & $3.0\times10^{-2}$ & m$^{2}$ \\
Volume, $Rw$ & $\mathcal{V}_{Rw}$        & $7.7\times10^{-5}$  & m$^{3}$ \\
Buoyancy mass   & $M_\mathrm{buoy}=\rho\,(\mathcal{V}_{Fw}+\mathcal{V}_{Rw})$
                           & $1.28$              & kg \\
Rider's height   & $h_{\text{rider}}$                    & $1.70$              & m \\
Mast length   & $\ell_\mathrm{mast}$   & $1.00$              & m \\
Feet distance & $\ell_\mathrm{feet}$      & $0.50$              & m \\
Distance $Fw$--$Rw$   & $\ell_{\text{wing,tot}}$ & $0.80$           & m \\
Distance $Fw$--$\mathbf{O} $ & $\ell_{Fw}$     & $0.15$              & m \\
Distance $Rw$--$\mathbf{O} $ & $\ell_{Rw}$     & $0.65$              & m \\
Moment of Inertia   & $I=m(l_{\text{mast}}+1/2h_{\text{rider}})^2$                    & $2.40\times10^{2}$  & kg\,m$^{2}$ \\
Pumping frequency   & $\omega = 2\pi\times1.65$ & $10.4$          & rad\,s$^{-1}$ \\
Rotational lift coefficient & $C_R$ & 1.1 & - \\
  \end{tabular}
\end{table}

\begin{figure}[!ht]
  \centering
  \includegraphics[width=0.9\textwidth]{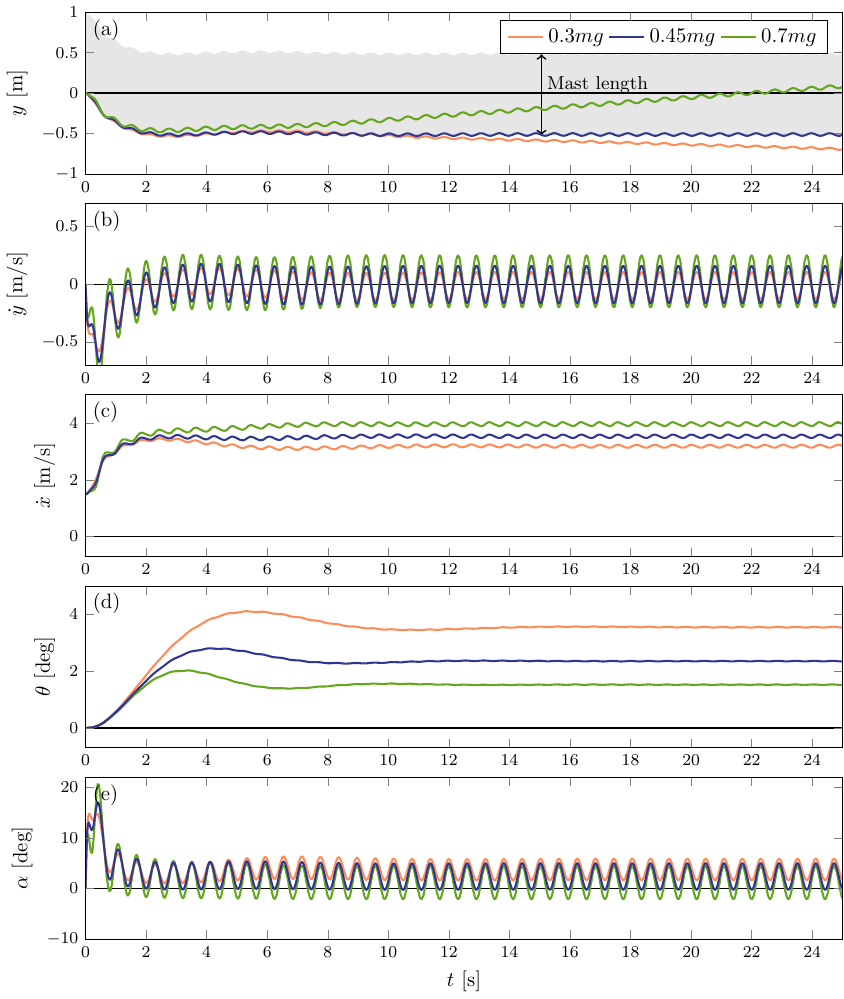}
  \caption{Solutions for three different pumping amplitudes $A_{\text{pump},{tot}}$ equal to $30\%$, $45\%$, and $70\%$ of body weight: (a) vertical position $y$ of the pivot point $\mathbf{O}$, (b) heave velocity $\dot{y}$, (c) forward velocity $\dot{x}$, (d) pitch angle $\theta$, and (e) angle of attack $\alpha$. Here, $k_\theta = A_{\text{pump},{tot}}/(1\,\mathrm{rad})$ is used.}
  \label{fig:Result}
\end{figure}
 
For the lift and drag coefficients, we use standard textbook empirical expressions for a foil, taken from \citet[Chapter 4]{anderson_cadou_foa7}:
\begin{align}
  C_D(\alpha) &\approx 0.2052\,\alpha^2 + 0.006, \\
  C_L(\alpha) &\approx 2\pi\,\alpha + 0.4, 
  \label{eq:liftcoef}
\end{align}
where $\alpha$ is expressed in radians. Note that the lift coefficient is non-zero at zero angle of attack due to the camber of the foil. These relations are valid in the range $\alpha \in [-10^\circ,\,15^\circ]$ \cite{anderson_cadou_foa7}. We set the rotational-lift coefficient to $C_R = 1.1$, following \citet{Li_Wang_Ristroph22}.

The coupled second-order ODEs (\ref{eq:finalx})--(\ref{eq:finalI}) are integrated numerically using the \texttt{ode45} solver in \textsc{Matlab}.  
 The initial horizontal velocity is set to $\dot{x}_0 = 1.5\ \mathrm{m/s}$, corresponding to the rider running with the board before entering the water, while all other initial values are set to zero. Again, for simplicity, we choose the pitch stiffness $k_\theta = A_{\text{pump},tot} / (1 \mathrm{rad})$, so that the pumping force amplitude $A_{\text{pump},{tot}}$ is the only control parameter of the model. If a different value of $k_\theta$ were chosen, one could simply retune $A_{\text{pump},{tot}}$ to recover a stable ride.

Figure \ref{fig:Result} shows the integration results as a function of time for three different pumping amplitudes $A_{\text{pump},{tot}}$ corresponding to $30\%$, $45\%$ and $70\%$ of the rider's body weight.
 In Figure \ref{fig:Result}a, the axial position of the pivot point is shown, and the mast length is also indicated as a shaded region. It shows that when the rider pumps too weakly, the rider sinks, whereas pumping too strongly causes the foil to leave the water, which may not be realistic. For the intermediate case, $A_{\text{pump,tot}} = 0.45 mg$, the rider can move forward stably, meaning the foil is pumped at 45\% of the rider’s maximum pumping capability ($mg$). The pivot point $\mathbf{O}$ oscillates about $50~\mathrm{cm}$ below the free surface, implying that the rider's board is about $50\ \mathrm{cm}$ above the surface, which is consistent with visual observations. The heaving velocity oscillates between $\pm 0.16~\mathrm{m/s}$ for this stable riding condition, and its amplitude increases when pumping harder.

The forward velocity is shown in Figure \ref{fig:Result}c, where it remains almost constant around $3.5~\mathrm{m/s}$ for the stable case. This value is about one order of magnitude larger (roughly $20$ times) than the heaving speed, which is also coherent with the observations. When pumping harder (more weakly), the forward velocity increases (decreases) slightly.

\begin{figure}[!ht]
  \centering
  \includegraphics[width=\textwidth]{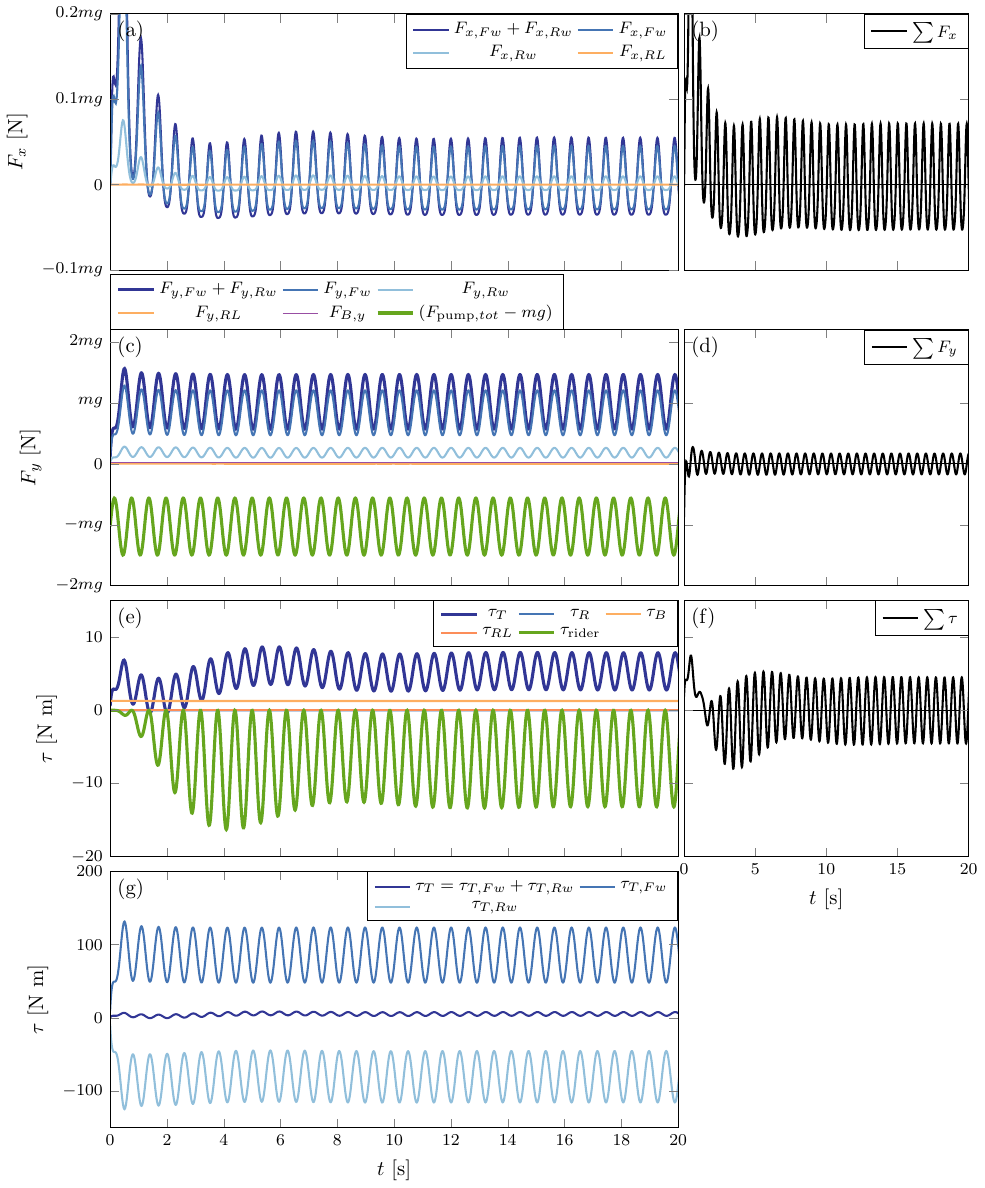}
  \caption{Detailed force and torque contributions for $A_{\text{pump},{tot}} = 0.45\,m g$ and $k_\theta = A_{\text{pump},{tot}}/(1\,\mathrm{rad})$. (a,b) forward thrusts, (c,d) axial forces, (e,f) torques, and (g) detailed view of the translational torque contributions from the front and rear wings ($Fw$ and $Rw$).}
  \label{fig:Result_detail}
\end{figure}

Figure \ref{fig:Result}d shows the pitch angle variation. Surprisingly little oscillation of the pitch angle is observed. For the optimal pumping case, the angle is almost constant at $\theta \simeq 0.04\  \mathrm{rad}$ (i.e. $2.3^\circ$) once stabilized. This justifies simplifying the model using a small angle approximation for $\theta$.

The angle of attack $\alpha$ is shown in Figure \ref{fig:Result}e, which varies from $0$ to $0.087\ \mathrm{rad}$ (about $5^\circ$). We confirm that this range remains within the validity of the linear lift-coefficient approximation in (\ref{eq:liftcoef}), which is assumed to hold for $\alpha \in [-10^\circ,\,15^\circ]$. For stronger pumping cases, $\alpha$ can reach slightly negative values, down to about $-2.1^\circ$, but thanks to the asymmetry of the foil, even at negative angle of attack it generates positive lift (the zero-lift angle is shifted to $\alpha \simeq 3.65^\circ$). 

Figure \ref{fig:Result_detail} shows the detailed force contribution of each directions for $A_{\text{pump,tot}} = 0.45 mg$ case. 
The contribution of each term to the forward (thrust) force is shown in Figure \ref{fig:Result_detail}ab. Once the motion is stabilised, the total thrust oscillates between approximately $-0.05\,mg$ and $0.07\,mg$. The dominant contribution comes from the aerodynamic lift-drag forces on the wings, with the front wing producing about five times more thrust than the rear wing. In contrast, the contribution from the rotation induced (Magnus-like) lift is very small. As seen from (\ref{eq:FyRL}), $F_{x,{RL}}$ is proportional to $\dot{\theta}\,\dot{y}$, which remains small for the present motion. Such a rotational lift force would only become significant for a body that rotates continuously without generating translational lift and drag, for example a rotating cylinder at $\alpha = 0$. In the pump foil configuration considered here, the rotational-lift contribution can therefore be neglected.
  
  The axial force contributions are shown in Figure \ref{fig:Result_detail}cd. There are more forces involved in the heaving motion. 
  First, it is interesting to note that the rider's weight and pumping force, $(F_{\text{pump}, tot} - mg)$, are mainly balanced by the aerodynamic lift-drag forces generated by the wings, $(F_{y,{Fw}} + F_{y,{Rw}})$. The other contributions, such as the rotational-lift force $F_{y,{RL}}$ (for the same reason discussed in the previous paragraph) and the buoyancy force $F_{b,y}$ (due to the small submerged volume and displaced water mass compared to the rider mass), are almost zero.
Comparing the contributions of the front and rear wings, the front wing produces more than four times the vertical force of the rear wing. This is mainly because its planform area is 4.6 times larger than that of the rear wing. 
As a result, the rider's weight and pumping forces, which oscillate between $-1.47\,mg$ and $-0.53\,mg$, are almost perfectly balanced by the lift generated by the wings, which ranges from $0.54\,mg$ to $1.43\,mg$. Consequently, the net vertical force $\sum F_y$ oscillates within approximately $\pm 0.17\,mg$ around zero.

Finally, the torque contributions are shown in Figure \ref{fig:Result_detail}ef. Due to the large lift generated by the front wing, the total hydrodynamic torque $\tau_T$ is predominantly positive, tending to rotate the foil in the nose-up direction. The buoyancy torque also contributes a positive moment. In contrast, the rider torque is always negative. This follows from our modelling assumption that only the rider's front leg contributes to the torque, 
 and that it performs only a ``pushing'' motion (negative torque) without any ``pulling'' motion (positive torque), as specified in (\ref{eq:taupump})-(\ref{eq:taupumpamp}). The resulting net torque $\sum \tau
$ oscillates between $-4.47$ and $4.45\ \text{N}\cdot\text{m}$.
It is interesting to decompose $\tau_T$ into the contributions from the front and rear wings. As discussed in the previous paragraph, the vertical force on the front wing, $F_{y,{Fw}}$, is about 4.6 times larger than that on the rear wing, $F_{y,{Rw}}$, mainly due to the larger planform area of the front wing, while their characteristic velocities are similar, \(V_{{Fw}} \simeq V_{{Rw}} \simeq \sqrt{\dot{x}^2+\dot{y}^2}\). Both forces generate a moment about the pivot: the front wing contribution is positive and the rear wing contribution is negative, as sketched in Figure \ref{fig:torque_T}. Although the front wing generates about 4.6 times more lift, its moment arm is much shorter, $l_{{Fw}} = 0.15\ \text{m}$ versus $l_{\mathrm{Rw}} = 0.65\ \text{m}$ (4.3 times shorter), so that the two torques become comparable in magnitude, as shown in Figure \ref{fig:Result_detail}g. The torque from the front wing oscillates between $47.0$ and $124.5\ \text{N}\cdot\text{m}$, whereas the rear wing torque ranges from $-116.6$ to $-44.3\ \text{N}\cdot\text{m}$. Both are roughly one order of magnitude larger than the resulting total torque, $\tau_T = \tau_{T,\mathrm{Fw}} + \tau_{T,\mathrm{Rw}}$. 
 This may explain why the rear wing is often referred to as the ``stabilizer'', as it stabilizes the overall torque but contribute only little on the overall lift.

Finally, we briefly comment on the role of pumping frequency (details in Appendix \ref{sec:depfreq}). At fixed total pumping force $F_{\mathrm{pump},{tot}} = 0.45\,mg$, the model predicts a finite band of frequencies, approximately $f \in [0.27,\,1.65]$ Hz, over which a stable submerged oscillation of the foil can be sustained with lower and equal pumping effort $\lesssim 0.45mg$. These trends provide a first indication of how sensitive the motion is to the choice of pumping frequency, but they should be viewed as qualitative.

\section{Conclusion and discussion}
In this study, we developed a minimal physical model that captures the essential dynamics of pump foil propulsion using a coupled second-order system of ordinary differential equations. The model is formulated for the combined rider-foil system: the rider is represented as a point mass coupled to a front and a rear wing, and the equations of motion are written about a pivot point located at the junction of the mast and the wing linker. Within this view point, the rider is not treated as an external load acting on the foil alone, but as part of the dynamical system itself, with the pumping action represented through an effective relative actuation contributing to both the translational forcing and the pitching torque.  Our formulation follows the spirit of the non-inertial models introduced for falling bodies in \citet{Li_Wang_Ristroph22,Pomerenk_Ristroph_2025}, but is adapted here to the pump foil configuration in an inertial (``observer-on-the-beach'') frame. The model includes quasi-steady lift and drag, buoyancy, rotational lift, and a nonlinear rotational drag term, together with a simplified description of rider control.

The most delicate and arguably most uncertain part of the model concerns the rider’s control. In order to keep the number of control parameters minimal, we made a set of deliberately strong but, we believe, reasonable first-order assumptions: the rear foot is fixed at the pivot and therefore does not generate torque; only the front foot contributes to the pitching torque; the total pumping force is prescribed as a sinusoid with fixed frequency and amplitude; the rider adjusts his center of mass to avoid a weight-induced torque; arm motion is neglected; and no explicit feedback control is modeled (the rider is represented as a prescribed actuator that depends only on the pitch angle and modulates only the pitching moment amplitude). These assumptions make the model tractable and transparent, and they are broadly consistent with qualitative observations of experienced riders, but they are clearly idealizations and should be challenged and refined in future work, using direct field measurements of actual riding strategies.

Within this set, the control space is reduced to a single dominant parameter, the total pumping force amplitude $A_{\text{pump},{tot}}$, with the frequency fixed as $1.65\ \mathrm{Hz}$ for a rider of mass $\sim 70\ \mathrm{kg}$. For this setup, we find that a pumping amplitude of order $45\%$ of the rider’s weight is sufficient to sustain a stable forward motion while keeping the pitch angle small throughout the cycle ($\sim 2.3^\circ$). 
However, in the field, sometimes we observe larger pitch-angle variations up to $\pm 10^\circ$ in the field. Our conjecture is that when the lake surface is calm (given that the rider is experienced one), the pitch angle remains close to zero, whereas under windy conditions the pitch angle varies more. It is not clear yet whether this is due to the surface waves impacting the lift and drag of the foil or to a destabilization of the rider’s equilibrium by the wind (or both).

A detailed force and moment balance shows that the rider’s effective weight and pumping force are primarily balanced by the lift generated by the front wing, while the rear wing contributes only about $22\%$ of the vertical support and is therefore almost negligible for overcoming the rider’s weight. However, the rear wing plays a crucial role in the torque balance: despite its smaller lift, its much larger moment arm (about $4.3$ times that of the front wing) allows it to counteract a large fraction of the nose-up moment generated by the front wing. The remaining torque is then regulated by the rider’s front leg, which supplies a nose-down torque. These results provide a physical picture of pump foil dynamics and clarify the distinct roles of the front wing, rear wing (stabiliser), and rider control.

Apart from demonstrating three representative pumping-force levels and  pumping frequency, we did not attempt a full parametric study. When one control parameter is changed, other parameters can generally be retuned to recover a stable riding condition, which limits the interpretive value of a broad parameter sweep at this stage. All forces in the model are expressed non-dimensionally with respect to the rider’s weight, so that changes in rider mass simply rescale the required pumping amplitude. Within a variation of $\pm 10 \mathrm{kg}$ in rider mass, we did not observe significant qualitative changes in the simulated dynamics.

Future work will focus on confronting and improving the present control assumptions using targeted measurements. In particular, field experiments with instrumented boards (separate force measurements under the front and rear feet, together with kinematic data for pitch and heave) are needed to determine the true time history and distribution of the pumping forces, and to assess whether the effective rider response is closer to open-loop or feedback control. In parallel, laboratory measurements of the hydrodynamic forces on realistic foil geometries will be used to refine the simplified lift and drag laws and to better quantify rotational effects. 
 Such measurements are important because the performance of unsteady propulsors can be highly sensitive to the details of the prescribed kinematics, as illustrated by \citet{Gehrke_2021}. 
The combined modelling and experimental efforts will be essential to move from the present minimal, strongly idealized control description towards a quantitatively predictive framework for pump foil performance and stability.

\appendix
\section{Added mass consideration} 
\label{sec:addedMass}
\begin{figure}[!ht]
  \centering
  \includegraphics[width=0.9\textwidth]{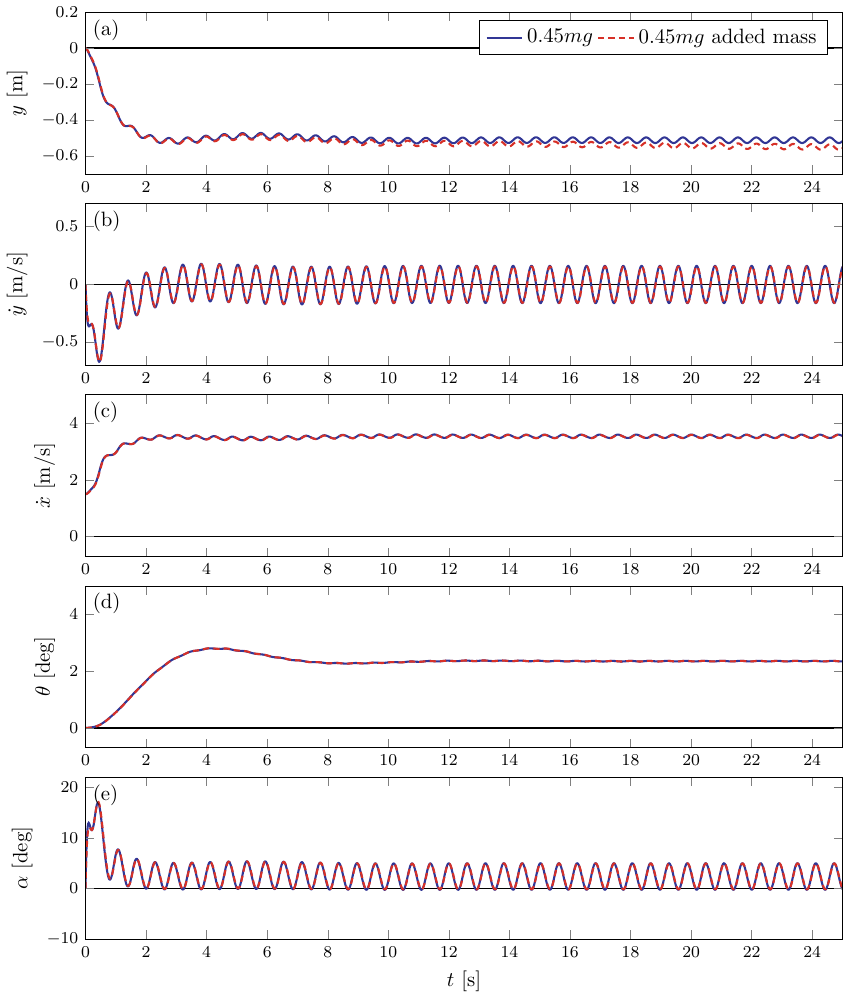}
  \caption{Solutions for added mass consideration solving equations (\ref{eq:added1})--(\ref{eq:added2}) for $A_{\text{pump},{tot}}=45mg$. After the calculation, the components are projected back to the inertial frame: (a) vertical position $y$ of the pivot point $\mathbf{O}$, (b) heave velocity $\dot{y}$, (c) forward velocity $\dot{x}$, (d) pitch angle $\theta$, and (e) angle of attack $\alpha$. Here, $k_\theta = A_{\text{pump},{tot}}/(1\,\mathrm{rad})$ is used.}
  \label{fig:Resultadded}
\end{figure}

 If we want to include added mass effect, it is much easier to put ourselves on the foil reference frame. Following the work of  \citet{Li_Wang_Ristroph22}: 
 \begin{align}
\dot{x}&=\dot{x}' \cos \theta-\dot{y}' \sin \theta, \label{eq:added1} \\
\dot{y}&=\dot{x}' \sin \theta+\dot{y}' \cos \theta, \\
\dot{\theta}&=\omega, \\
\left(m+m_{11}\right) \ddot{x}'&=\left(m+m_{22}\right) \dot{\theta} \dot{y}'-(m_{22,Fw} l_{Fw}+m_{22,Rw} l_{Rw})\dot{\theta}^2  +F_{x^{\prime}}-m^{\prime} g \sin \theta, \\
\left(m+m_{22}\right) \ddot{y}'&=-\left(m+m_{11}\right) \dot{\theta} \dot{x}'+(m_{22,Fw}l_{Fw}+m_{22,Rw} l_{Rw}) \ddot{\theta} +F_{y^{\prime}}-m^{\prime} g \cos \theta, \\
\left(I+I_a\right) \ddot{\theta}&=\tau_{\text{rider}} + \tau_B + \tau_T + \tau_R +\tau_{RL}.\label{eq:added2}
\end{align} 
As we work in a non-inertial frame, fictitious forces appear. The added masses in the $x'$ and $y'$ directions are denoted $m_{11}$ and $m_{22}$, respectively. We adopt a simple thin-plate added mass model, following \citet{Li_Wang_Ristroph22}:
\begin{align}
m_{11} &= 0, \\
m_{22} &= m_{22,Fw} + m_{22,Rw} \\
&= \frac{\pi}{4} \rho \left( c_{Fw}^2 b_{Fw} + c_{Rw}^2 b_{Rw} \right).
\end{align}
The corresponding added moment of inertia $I_a$ is approximated by an $mr^2$ contribution:
\begin{equation}
I_a = m_{22,Fw} l_{Fw}^2 + m_{22,Rw} l_{Rw}^2.
\end{equation}
Numerically, we obtain $m_{22} \simeq 14.4 + 1.4 = 15.8~\mathrm{kg}$ and $I_a \simeq 0.92~\mathrm{kg\,m^2}$, which correspond to about $20\%$ of the rider’s mass and $0.4\%$ of the total moment of inertia of the system.
Figure \ref{fig:Resultadded} compares the results obtained with and without added mass. As shown, the influence of added mass on the dynamics is negligible.

\section{Dependence on pumping frequency} 
\label{sec:depfreq}
\begin{figure}[!ht]
  \centering
  \includegraphics[width=0.9\textwidth]{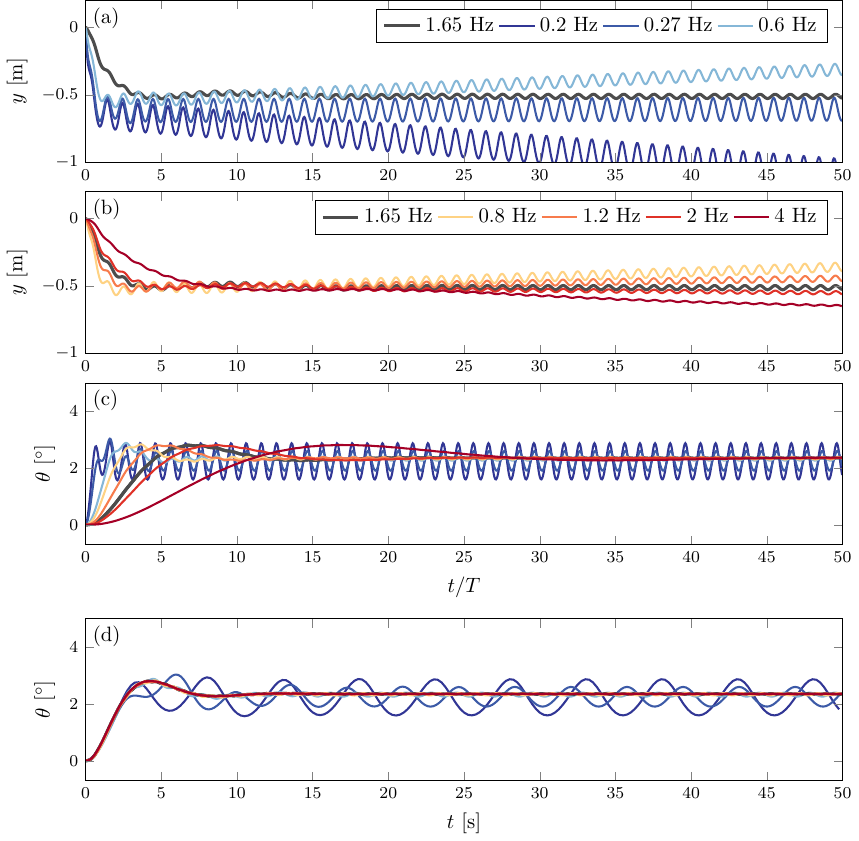}
  \caption{(a,b) Foil vertical position $y$ and (c,d) pitch angle $\theta$ for different pumping frequencies, compared to the reference case at $f = 1.65$ Hz with $F_{\mathrm{pump},{tot}} = 0.45\,mg$. In (a--c), time is normalized by the pumping period $T = 1/f$ and 50 cycles are shown. In (d), time is dimensional.}
  \label{fig:dependancyFreq}
\end{figure}

Figure \ref{fig:dependancyFreq} shows how the motion depends on the pumping frequency. All parameters are kept at their default values, and we vary only the pumping frequency while fixing the total pumping force at $F_{\mathrm{pump},{tot}} = 0.45\,mg$. Results are shown for $f = 0.2$, $0.27$, $0.6$, $0.8$, $1.2$, $2$, and $4$ Hz. In Figs. \ref{fig:dependancyFreq}(a)-(c), time is nondimensionalized by the pumping period $T = 1/f$, so that $t/T$ corresponds to the number of cycles, and we display 50 successive periods. For simplicity, we show only the foil vertical position $y$ and the pitch angle $\theta$.

Two main trends can be identified when examining the vertical position $y$ in Figs. \ref{fig:dependancyFreq}(a,b). First, the oscillation amplitude of $y$ decreases as the pumping frequency increases. Second, the mean vertical position of the foil and the qualitative nature of the motion vary in a non-monotonic way with $f$. At very low frequency, $f = 0.2$ Hz, the rider gradually sinks. At $f = 0.27$Hz, the motion becomes approximately periodic and the foil remains close to its initial submerged level. For a slightly higher frequency, $f = 0.5$ Hz, the oscillations grow large enough for the foil to periodically exit the water. As the frequency is further increased, the foil continues to approach or intersect the free surface until about $f = 1.65$ Hz, where the model again predicts a stable, fully submerged oscillation. For frequencies higher than $1.65$ Hz and at fixed $F_{\mathrm{pump},{tot}} = 0.45\,mg$, the pumping becomes insufficient and the rider eventually sinks.

These results indicate that, for a given rider and geometry, there exists only a finite range of pumping frequencies over which a moderate pumping force (here $45\%$ of the rider’s weight) can sustain a stable underwater ride. Within the approximate range $f \in [0.27,\,1.65]$ Hz, one could in principle reduce the pumping force slightly below $0.45\,mg$ and still maintain a periodic motion, thereby saving some effort. Outside this range, $f < 0.27$ Hz or $f > 1.65$ Hz, a stronger pumping force would be required to avoid sinking.

The behaviour of the pitch angle is shown in Figs. \ref{fig:dependancyFreq}(c,d). When plotted against the nondimensional time $t/T$ (Fig. \ref{fig:dependancyFreq}(c)), all cases relax towards small-amplitude oscillations of order $2^{\circ}$, with larger oscillation amplitudes at lower pumping frequencies. 
When plotted against the dimensional time $t$ (Fig. \ref{fig:dependancyFreq}(d)), the initial transient growth of $\theta$ is nearly linear and exhibits an almost identical slope for all frequencies, indicating that the early-time pitch dynamics are  insensitive to the choice of $f$ and are instead controlled by the common torque and inertia scales of the system. This behaviour is reflects that an existence of a characteristic pitch-adjustment timescale, set by the ratio of the total moment of inertia to the typical hydrodynamic and pumping torques, which governs the transient response before frequency dependent effects become important.

We finally emphasize that these frequency dependent trends are predictions of an idealized model and should be interpreted with caution. In particular, the present description neglects several sources of complexity such as unsteady three-dimensional flow effects, wave generation and breaking at the free surface, and variations in the rider’s actual control strategy. Quantitative bounds on the stable-frequency range and the required pumping effort will therefore need to be validated against systematic measurements.

\section*{Acknowledgements}
The authors would like to thank A. Blanchard from the Laidlaw programme, as well as B. Lupton, E. Prigent, I. Tavier and L. Perata, the students who worked on the \emph{Projet d'ingénierie simultanée} 2025. We are also grateful to the pump foil rider I. Labarthe for sharing his experience on the water. We also thank T. Schneider, K. Mulleners and J.-P. Boucher for fruitful discussions. 
\bibliography{biblio_all.bib}

\end{document}